\documentclass[conference, letterpaper]{IEEEtran}
\IEEEoverridecommandlockouts

\usepackage[hyphens,spaces,obeyspaces]{url}
\usepackage[breaklinks=true]{hyperref}
\usepackage{algorithm}
\usepackage{algpseudocode}
\usepackage[usenames]{color}
\usepackage[table,xcdraw]{xcolor} 
\usepackage{array}
\usepackage[most]{tcolorbox}
\usepackage{subfig}
\usepackage{float}
\usepackage{multicol}
\usepackage{caption}
\usepackage{tikz}
\usepackage{color}
\usepackage{gensymb}
\usepackage{cite}
\usepackage{graphicx}
\usepackage{psfrag} 
\usepackage{paralist}
\usepackage{multirow}
\usepackage[acronym]{glossaries}
\usepackage{amsmath,stackrel}
\usepackage{amsfonts}
\usepackage{mathalfa}
\usepackage{amssymb,mathtools}
\usepackage{amscd}
\usepackage{tikz}
\usepackage{verbatim}
\usepackage{enumerate}                     
\usepackage{amsthm}
\usepackage{eucal}
\usepackage[nolist]{acronym}
\usepackage[normalem]{ulem}
\usepackage{fancyhdr}
\usepackage{pgfplots}
\usepackage{pgfplotstable}
\usepackage{epsfig}
\usepackage{cite}
\usepackage{textcomp}
\usepackage{psfrag}
\usepackage{arydshln}
\usepackage{url}
\usepackage{soul}
\usepackage{comment}
\usepackage{hyperref}
\usepackage{graphicx,color}
\usepackage{epstopdf}
\usepackage[nolist]{acronym}
\usepackage{url}
\usepackage[capitalise]{cleveref}
\usepackage{pgfplots}
\usepackage{pgfplotstable}
\usepackage{nopageno}

\def\BibTeX{{\rm B\kern-.05em{\sc i\kern-.025em b}\kern-.08em
    T\kern-.1667em\lower.7ex\hbox{E}\kern-.125emX}}
    
\begin{document}

\title{Remote Breathing Monitoring Using LiDAR Technology\\ \vspace{-4 mm}
}

\author{\IEEEauthorblockN{Omar Rinchi}
\IEEEauthorblockA{
Missouri University of\\ Science and Technology\\
Rolla, MO, USA\\
Email: omar.rinchi@mst.edu}
\and
\IEEEauthorblockN{Ahmad Alsharoa}
\IEEEauthorblockA{
Missouri University of\\ Science and Technology\\
Rolla, MO, USA\\
Email: aalsharoa@mst.edu}
\and
\IEEEauthorblockN{Denise A. Baker}
\IEEEauthorblockA{
Missouri University of \\ Science and Technology\\
Rolla, MO, USA\\
Email: bakerden@mst.edu}} 

\maketitle

\def\scheme{\textit{\ac{SIDCo} }} 
\def\d{d} 
\def\k{k} 
 \newcommand{\topk}{{\text{Top}_{\k}}} 
 \newcommand{\randk}{{\text{Rand}_{\k}}} 

\def\N{N}  
\def\x{\mathbf x}
\def\xi [#1]{\x_{\{#1\}}}
\def\xin[#1]{{\x}_{\{#1\}}^{n}}
\def\gr{g} 
\def\g{\mathbf \gr}
\def\gin [#1]{{\g}_{#1}^{n}}
\def\G{G} 
\def\C{\mathbb{C}_{\k}} 
\def\Tk[#1]{{\mathbb{T}}_{\k}\left\{ #1 \right\} } 
\def\Ceta{\mathbb{C}_{\eta}} 
\def\sigmak{\sigma_{\k}}  
\def\p{p} 
\def\b{\beta} 
\def\a{\alpha} 
\def\loc{a}
\def\bhat{\hat{\beta}} 
\def\ahat{\hat{\alpha}} 
\def\lochat{\hat{a}}
\def\Hbr{\mathbf{H}^{\text{BR}}\in \mathbb{C}^{N^\text{B} \times N^{\text{R}}}}
\def\Aphibr{{\textbf{a}(\boldsymbol{\phi}^\text{BR},\textbf{ r}^\text{BR})}\in \mathbb{C}^{N^\text{B} }}
\def\Athetabr{\textbf{a}(\boldsymbol{\theta}^\text{BR},\textbf{d}^\text{BR})\in\mathbb{C}^{N^\text{R} }}
\def\Hru{\mathbf{H}^{\text{RU}}\in \mathbb{C}^{N^\text{R} \times N^{\text{U}}}}

\def\H{\mathbf{H}\in \mathbb{C}^{N^\text{B} \times N^{\text{U}}}}

\def\sign{\textrm{sign}}                                              
\def\erf{\textrm{erf}}
\def\erfc{\textrm{erfc}}

\def\Hzero{{{\mathcal{H}}_{0}}}
\def\Hone{{{\mathcal{H}}_{1}}}
\def\Hl{{{\mathcal{H}}_{l}}}

\def\Dzero{{{\mathcal{D}}_{0}}}
\def\Done{{{\mathcal{D}}_{1}}}

\def\testH1overH0{\begin{array}{l}
		{\mathcal{H}}_1 \\
		\gtrless \\
		{\mathcal{H}}_0
\end{array}}

\def\testH0overH1{\begin{array}{l}
		{\mathcal{H}}_0 \\
		\gtrless \\
		{\mathcal{H}}_1
\end{array}}

\def\testD1overD0{{ \begin{array}{l}
			{\mathcal{D}}_1 \\
			\gtrless \\
			{\mathcal{D}}_0
\end{array}}}

\begin{acronym}

\acro{5G-NR}{5G New Radio}
\acro{3GPP}{3rd Generation Partnership Project}
\acro{AC}{address coding}
\acro{ACF}{autocorrelation function}
\acro{ACR}{autocorrelation receiver}
\acro{ADC}{analog-to-digital converter}
\acrodef{aic}[AIC]{Analog-to-Information Converter}     
\acro{AIC}[AIC]{Akaike information criterion}
\acro{aric}[ARIC]{asymmetric restricted isometry constant}
\acro{arip}[ARIP]{asymmetric restricted isometry property}

\acro{ARQ}{automatic repeat request}
\acro{AUB}{asymptotic union bound}
\acrodef{awgn}[AWGN]{Additive White Gaussian Noise}     
\acro{AWGN}{additive white Gaussian noise}
\acro{CRLB}{Cramér-Rao lower bound} 
\acro{RSS}{received signal strength} 

\acro{APSK}[PSK]{asymmetric PSK} 

\acro{waric}[AWRICs]{asymmetric weak restricted isometry constants}
\acro{warip}[AWRIP]{asymmetric weak restricted isometry property}
\acro{BCH}{Bose, Chaudhuri, and Hocquenghem}        
\acro{BCHC}[BCHSC]{BCH based source coding}
\acro{BEP}{bit error probability}
\acro{BFC}{block fading channel}
\acro{BG}[BG]{Bernoulli-Gaussian}
\acro{BGG}{Bernoulli-Generalized Gaussian}
\acro{BPAM}{binary pulse amplitude modulation}
\acro{BPDN}{Basis Pursuit Denoising}
\acro{BPPM}{binary pulse position modulation}
\acro{BPSK}{binary phase shift keying}
\acro{BPZF}{bandpass zonal filter}
\acro{BSC}{binary symmetric channels}              
\acro{BU}[BU]{Bernoulli-uniform}
\acro{BER}{bit error rate}
\acro{BS}{base station}

\acro{CP}{Cyclic Prefix}
\acrodef{cdf}[CDF]{cumulative distribution function}   
\acro{CDF}{cumulative distribution function}
\acrodef{c.d.f.}[CDF]{cumulative distribution function}
\acro{CCDF}{complementary cumulative distribution function}
\acrodef{ccdf}[CCDF]{complementary CDF}               
\acrodef{c.c.d.f.}[CCDF]{complementary cumulative distribution function}
\acro{CD}{cooperative diversity}

\acro{CDMA}{Code Division Multiple Access}
\acro{ch.f.}{characteristic function}
\acro{CIR}{channel impulse response}
\acro{cosamp}[CoSaMP]{compressive sampling matching pursuit}
\acro{CR}{cognitive radio}
\acro{cs}[CS]{compressed sensing}                   
\acrodef{cscapital}[CS]{Compressed sensing} 
\acrodef{CS}[CS]{compressed sensing}
\acro{CSI}{channel state information}
\acro{CCSDS}{consultative committee for space data systems}
\acro{CC}{convolutional coding}
\acro{Covid19}[COVID-19]{Coronavirus disease}

\acro{DAA}{detect and avoid}
\acro{DAB}{digital audio broadcasting}
\acro{DCT}{discrete cosine transform}
\acro{dft}[DFT]{discrete Fourier transform}
\acro{DR}{distortion-rate}
\acro{DS}{direct sequence}
\acro{DS-SS}{direct-sequence spread-spectrum}
\acro{DTR}{differential transmitted-reference}
\acro{DVB-H}{digital video broadcasting\,--\,handheld}
\acro{DVB-T}{digital video broadcasting\,--\,terrestrial}
\acro{DL}{downlink}
\acro{DSSS}{Direct Sequence Spread Spectrum}
\acro{DFT-s-OFDM}{Discrete Fourier Transform-spread-Orthogonal Frequency Division Multiplexing}
\acro{DAS}{distributed antenna system}
\acro{DNA}{Deoxyribonucleic Acid}

\acro{EC}{European Commission}
\acro{EED}[EED]{exact eigenvalues distribution}
\acro{EIRP}{Equivalent Isotropically Radiated Power}
\acro{ELP}{equivalent low-pass}
\acro{eMBB}{enhanced mobile broadband}
\acro{EMF}{electric and magnetic fields}
\acro{EU}{European union}

\acro{FC}[FC]{fusion center}
\acro{FCC}{Federal Communications Commission}
\acro{FEC}{forward error correction}
\acro{FFT}{fast Fourier transform}
\acro{FH}{frequency-hopping}
\acro{FH-SS}{frequency-hopping spread-spectrum}
\acrodef{FS}{Frame synchronization}
\acro{FSsmall}[FS]{frame synchronization}  
\acro{FDMA}{Frequency Division Multiple Access}

\acro{GA}{Gaussian approximation}
\acro{GF}{Galois field }
\acro{GG}{Generalized-Gaussian}
\acro{GIC}[GIC]{generalized information criterion}
\acro{GLRT}{generalized likelihood ratio test}
\acro{GPS}{Global Positioning System}
\acro{GMSK}{Gaussian minimum shift keying}
\acro{GSMA}{Global System for Mobile communications Association}

\acro{HAP}{high altitude platform}

\acro{IDR}{information distortion-rate}
\acro{IFFT}{inverse fast Fourier transform}
\acro{iht}[IHT]{iterative hard thresholding}
\acro{i.i.d.}{independent, identically distributed}
\acro{IoT}{internet of things}                      
\acro{IR}{impulse radio}
\acro{lric}[LRIC]{lower restricted isometry constant}
\acro{lrict}[LRICt]{lower restricted isometry constant threshold}
\acro{ISI}{intersymbol interference}
\acro{ITU}{International Telecommunication Union}
\acro{ICNIRP}{International Commission on Non-Ionizing Radiation Protection}
\acro{IEEE}{Institute of Electrical and Electronics Engineers}
\acro{ICES}{IEEE international committee on electromagnetic safety}
\acro{IEC}{International Electrotechnical Commission}
\acro{IARC}{International Agency on Research on Cancer}
\acro{IS-95}{Interim Standard 95}

\acro{LEO}{low earth orbit}
\acro{LF}{likelihood function}
\acro{LLF}{log-likelihood function}
\acro{LLR}{log-likelihood ratio}
\acro{LLRT}{log-likelihood ratio test}
\acro{LoS}{line-of-sight}
\acro{LRT}{likelihood ratio test}
\acro{wlric}[LWRIC]{lower weak restricted isometry constant}
\acro{wlrict}[LWRICt]{LWRIC threshold}
\acro{LPWAN}{low power wide area network}
\acro{LoRaWAN}{Low power long Range Wide Area Network}
\acro{NLOS}{non-line-of-sight}

\acro{MB}{multiband}
\acro{MC}{multicarrier}
\acro{MDS}{mixed distributed source}
\acro{MF}{matched filter}
\acro{m.g.f.}{moment generating function}
\acro{MI}{mutual information}
\acro{MIMO}{multiple-input multiple-output}
\acro{MISO}{multiple-input single-output}
\acrodef{maxs}[MJSO]{maximum joint support cardinality}                       
\acro{ML}[ML]{maximum likelihood}
\acro{MMSE}{minimum mean-square error}
\acro{MMV}{multiple measurement vector}
\acrodef{MOS}{model order selection}
\acro{M-PSK}[${M}$-PSK]{$M$-ary phase shift keying}                       
\acro{M-APSK}[${M}$-PSK]{$M$-ary asymmetric PSK} 

\acro{M-QAM}[$M$-QAM]{$M$-ary quadrature amplitude modulation}
\acro{MRC}{maximal ratio combiner}                  
\acro{maxs}[MSO]{maximum sparsity order}                                      
\acro{M2M}{machine to machine}                                                
\acro{MUI}{multi-user interference}
\acro{mMTC}{massive machine type communications}      
\acro{mm-wave}{millimeter-wave}
\acro{MP}{mobile phone}
\acro{MPE}{maximum permissible exposure}
\acro{MAC}{media access control}
\acro{NB}{narrowband}
\acro{NBI}{narrowband interference}
\acro{NLA}{nonlinear sparse approximation}
\acro{NLOS}{Non-Line of Sight}
\acro{NTIA}{National Telecommunications and Information Administration}
\acro{NTP}{National Toxicology Program}
\acro{NHS}{National Health Service}

\acro{OC}{optimum combining}                             
\acro{OC}{optimum combining}
\acro{ODE}{operational distortion-energy}
\acro{ODR}{operational distortion-rate}
\acro{OFDM}{orthogonal frequency-division multiplexing}
\acro{omp}[OMP]{orthogonal matching pursuit}
\acro{OSMP}[OSMP]{orthogonal subspace matching pursuit}
\acro{OQAM}{offset quadrature amplitude modulation}
\acro{OQPSK}{offset QPSK}
\acro{OFDMA}{Orthogonal Frequency-division Multiple Access}
\acro{OPEX}{Operating Expenditures}
\acro{OQPSK/PM}{OQPSK with phase modulation}

\acro{PAM}{pulse amplitude modulation}
\acro{PAR}{peak-to-average ratio}
\acrodef{pdf}[PDF]{probability density function}                      
\acro{PDF}{probability density function}
\acrodef{p.d.f.}[PDF]{probability distribution function}
\acro{PDP}{power dispersion profile}
\acro{PMF}{probability mass function}                             
\acrodef{p.m.f.}[PMF]{probability mass function}
\acro{PN}{pseudo-noise}
\acro{PPM}{pulse position modulation}
\acro{PRake}{Partial Rake}
\acro{PSD}{power spectral density}
\acro{PSEP}{pairwise synchronization error probability}
\acro{PSK}{phase shift keying}
\acro{PD}{power density}
\acro{8-PSK}[$8$-PSK]{$8$-phase shift keying}
\acro{PRS}{positioning reference signal}

\acro{FSK}{frequency shift keying}

\acro{QAM}{Quadrature Amplitude Modulation}
\acro{QPSK}{quadrature phase shift keying}
\acro{OQPSK/PM}{OQPSK with phase modulator }

\acro{RD}[RD]{raw data}
\acro{RDL}{"random data limit"}
\acro{ric}[RIC]{restricted isometry constant}
\acro{rict}[RICt]{restricted isometry constant threshold}
\acro{rip}[RIP]{restricted isometry property}
\acro{ROC}{receiver operating characteristic}
\acro{rq}[RQ]{Raleigh quotient}
\acro{RS}[RS]{Reed-Solomon}
\acro{RSC}[RSSC]{RS based source coding}
\acro{r.v.}{random variable}                               
\acro{R.V.}{random vector}
\acro{RMS}{root mean square}
\acro{RFR}{radiofrequency radiation}
\acro{RIS}{reconfigurable intelligent surface}
\acro{RNA}{RiboNucleic Acid}

\acro{SA}[SA-Music]{subspace-augmented MUSIC with OSMP}
\acro{SCBSES}[SCBSES]{Source Compression Based Syndrome Encoding Scheme}
\acro{SCM}{sample covariance matrix}
\acro{SEP}{symbol error probability}
\acro{SG}[SG]{sparse-land Gaussian model}
\acro{SIMO}{single-input multiple-output}
\acro{SINR}{signal-to-interference plus noise ratio}
\acro{SIR}{signal-to-interference ratio}
\acro{SISO}{single-input single-output}
\acro{SMV}{single measurement vector}
\acro{SNR}[\textrm{SNR}]{signal-to-noise ratio} 
\acro{sp}[SP]{subspace pursuit}
\acro{SS}{spread spectrum}
\acro{SW}{sync word}
\acro{SAR}{specific absorption rate}
\acro{SSB}{synchronization signal block}

\acro{TH}{time-hopping}
\acro{ToA}{time-of-arrival}
\acro{TR}{transmitted-reference}
\acro{TW}{Tracy-Widom}
\acro{TWDT}{TW Distribution Tail}
\acro{TCM}{trellis coded modulation}
\acro{TDD}{time-division duplexing}
\acro{TDMA}{Time Division Multiple Access}

\acro{UAV}{unmanned aerial vehicle}
\acro{uric}[URIC]{upper restricted isometry constant}
\acro{urict}[URICt]{upper restricted isometry constant threshold}
\acro{UWB}{ultrawide band}
\acro{UWBcap}[UWB]{Ultrawide band}   
\acro{URLLC}{ultra reliable low latency communications}
         
\acro{wuric}[UWRIC]{upper weak restricted isometry constant}
\acro{wurict}[UWRICt]{UWRIC threshold}                
\acro{UE}{user equipment}
\acro{UL}{uplink}

\acro{WiM}[WiM]{weigh-in-motion}
\acro{WLAN}{wireless local area network}
\acro{wm}[WM]{Wishart matrix}                               
\acroplural{wm}[WM]{Wishart matrices}
\acro{WMAN}{wireless metropolitan area network}
\acro{WPAN}{wireless personal area network}
\acro{wric}[WRIC]{weak restricted isometry constant}
\acro{wrict}[WRICt]{weak restricted isometry constant thresholds}
\acro{wrip}[WRIP]{weak restricted isometry property}
\acro{WSN}{wireless sensor network}                        
\acro{WSS}{wide-sense stationary}
\acro{WHO}{World Health Organization}
\acro{Wi-Fi}{wireless fidelity}

\acro{sss}[SpaSoSEnc]{sparse source syndrome encoding}

\acro{VLC}{visible light communication}
\acro{VPN}{virtual private network} 
\acro{RF}{radio frequency}
\acro{FSO}{free space optics}
\acro{IoST}{Internet of space things}

\acro{GSM}{Global System for Mobile Communications}
\acro{2G}{second-generation cellular network}
\acro{3G}{third-generation cellular network}
\acro{4G}{fourth-generation cellular network}
\acro{5G}{5th-generation cellular network}	
\acro{gNB}{next generation node B base station}
\acro{NR}{New Radio}
\acro{UMTS}{Universal Mobile Telecommunications Service}
\acro{LTE}{Long Term Evolution}

\acro{QoS}{quality of service}

\acro{ULA}{uniform linear array}
\acro{MS}{mobile station}
\acro{6G}{sixth-generation cellular networks}
\acro{EIV}{error-in-variable model}
\acro{DoA}{direction of arrival}
\acro{AoA}{angle of arrival}
\acro{AoD}{angle of departure}
\acro{LS}{least square}
\acro{SLAM}{simultaneous localization and mapping}
\acro{M-OMP}{multiple orthogonal matching pursuit}
\acro{JLZA}{joint $\ell_{2,0}$ approximation}
\acro{T-MSBL}{temporal multiple sparse Bayesian learning}
\acro{MAP}{maximum a posteriori}
\acro{ToA}{time of arrival}
\acro{PSO}{particle swarm optimization}
\acro{SDP}{semidefinite programming}
\acro{ES}{exhaustive search}
\acro{ISLAC}{integrated sensing, localization, and communication}
\acro{SDS}{software defined surfaces}
\acro{LiDAR}{light detection and ranging}
\acro{3D}{three dimensional}
\acro{ROI}{region of interest}
\acro{DBSCAN}{Density-based spatial clustering of applications with noise}
\acro{2D}{two dimensional}
\acro{RNN}{recurrent neural network}
\acro{LSTM}{long-sort term memory}
\acro{GRU}{gated recurrent unit}
\acro{SCR}{static clutter removal}
\acro{ABG}{arterial blood gas}
\acro{PII}{personal identifi able information}
\acro{PAR}{patients' activity recognition}

\end{acronym}

\begin{abstract}
Breathing monitoring is crucial in healthcare for early detection of health issues, but traditional methods face challenges like invasiveness, privacy concerns, and limited applicability in daily settings. This paper introduces \ac{LiDAR} sensors as a remote, privacy-respecting alternative for monitoring breathing metrics, including inhalation/exhalation patterns, respiratory rates, breath depth, and detecting breathlessness. We highlight \ac{LiDAR}'s ability to function across various postures, presenting empirical evidence of its accuracy and reliability. Our findings position \ac{LiDAR} as an innovative solution in breathing monitoring, offering significant advantages over conventional methods.

\end{abstract}

\begin{IEEEkeywords}
light detection and ranging (LiDAR),  apeana, breathing, healthcare, and respiratory rate. \vspace{-3 mm}
\end{IEEEkeywords}

\section{Introduction}
Breathing monitoring emerges as a powerful tool for early health detection and management, capturing essential metrics like respiratory rate and tidal volume. It shines in everyday settings, alerting to anomalies that may indicate health issues, from sleep apnea and stress to potential respiratory infections \cite{2022_Bahrami}. By offering real-time insights, it empowers individuals to act swiftly, fostering proactive health practices. This not only transforms medical care but also enhances daily well-being, making a significant impact on life quality.

Breathing monitoring technologies fall into invasive, non-invasive, and remote categories. Invasive methods, like the \ac{ABG} analysis and tracheal intubation, offer high precision but are uncomfortable and restricted to medical settings \cite{2022_Costanzo}. Non-invasive techniques use external devices like chest straps and smart textiles \cite{2019_Massaroni}, which are user-friendly but can be cumbersome and socially awkward for continuous wear \cite{2020_Baker}. Remote monitoring, utilizing technologies like thermal and RGB-D depth cameras, captures breathing data without physical contact, offering a less intrusive option for continuous observation \cite{2017_Prochazka, 2021_Massaroni}.

Remote breathing monitoring, while advantageous for its passive, non-intrusive tracking and comfort, faces significant challenges that impact its efficacy. Environmental factors like ambient temperature and lighting conditions critically influence the accuracy of technologies such as thermal and RGB-D sensors. For example, thermal cameras struggle with dynamic activities due to temperature variability \cite{2014_James}, and RGB-D sensors perform poorly in low light or direct sunlight. Moreover, these technologies face limitations in detecting dark colors, potentially leading to biased outcomes against individuals with darker skin tones \cite{2010_GameSpot, 2022_Angwin}. Privacy concerns also emerge with the use of video and thermal cameras, posing ethical considerations \cite{2020_EFF}. 

We present \ac{LiDAR} technology as a robust solution to traditional breathing monitoring challenges, leveraging laser light for remote sensing and generating 3D mappings without being influenced by ambient temperature, lighting, or object color and texture. This color-blind approach avoids potential color discrimination and enhances privacy by minimizing the risk of identifying personal features through sparse \ac{LiDAR}-generated point clouds \cite{2023_Rinchi}. \ac{LiDAR}'s versatility is demonstrated in healthcare applications, from activity recognition using deep-learning \cite{2023_Rinchi_B} to estimating tidal volume in patients \cite{2021_Hill, 2022_Bin}, showcasing its potential for remote, privacy-preserving monitoring in diverse settings.

This study explores \ac{LiDAR}'s capability to monitor various respiratory metrics, extending beyond the tidal volume estimation in supine subjects as seen in previous research \cite{2021_Hill, 2022_Bin}. We broaden our investigation by $(i)$ utilizing \ac{LiDAR} for detailed breathing analysis, including tracking inhalation/exhalation patterns, estimating respiratory rate, assessing breath depth, and detecting breathlessness episodes; and $(ii)$ showcasing \ac{LiDAR}'s adaptability in monitoring breathing across different postures, not limited to lying down positions.

\begin{figure}[t!]
         \centering
         \includegraphics[width=3in]{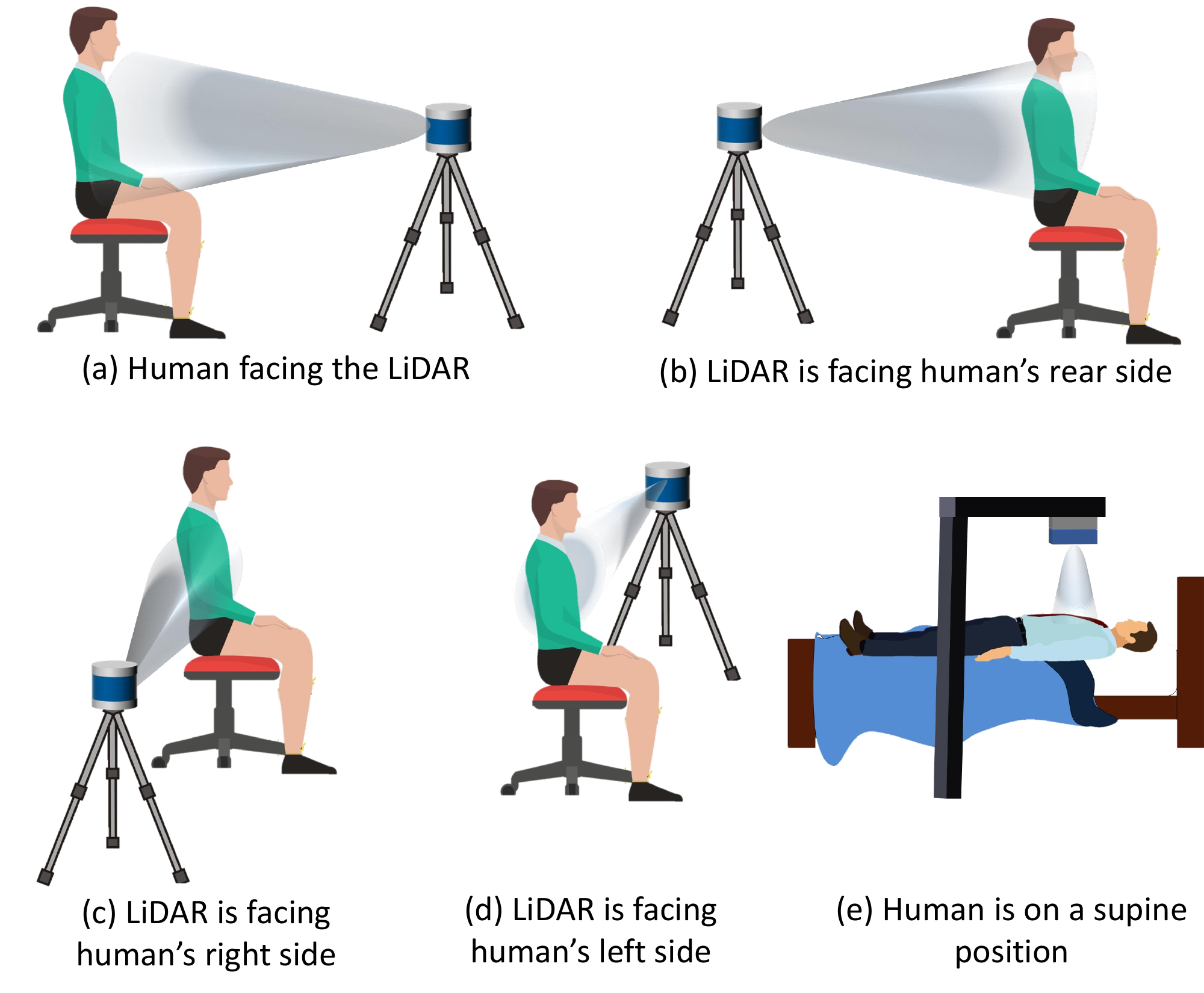}
         \vspace{-2 mm}
         \caption{The proposed system scenario and experimental setup.}
         \label{Fig1}
         \vspace{-7 mm}
\end{figure}

\begin{figure*}[t!]
         \centering
         \includegraphics[width=4.9in]{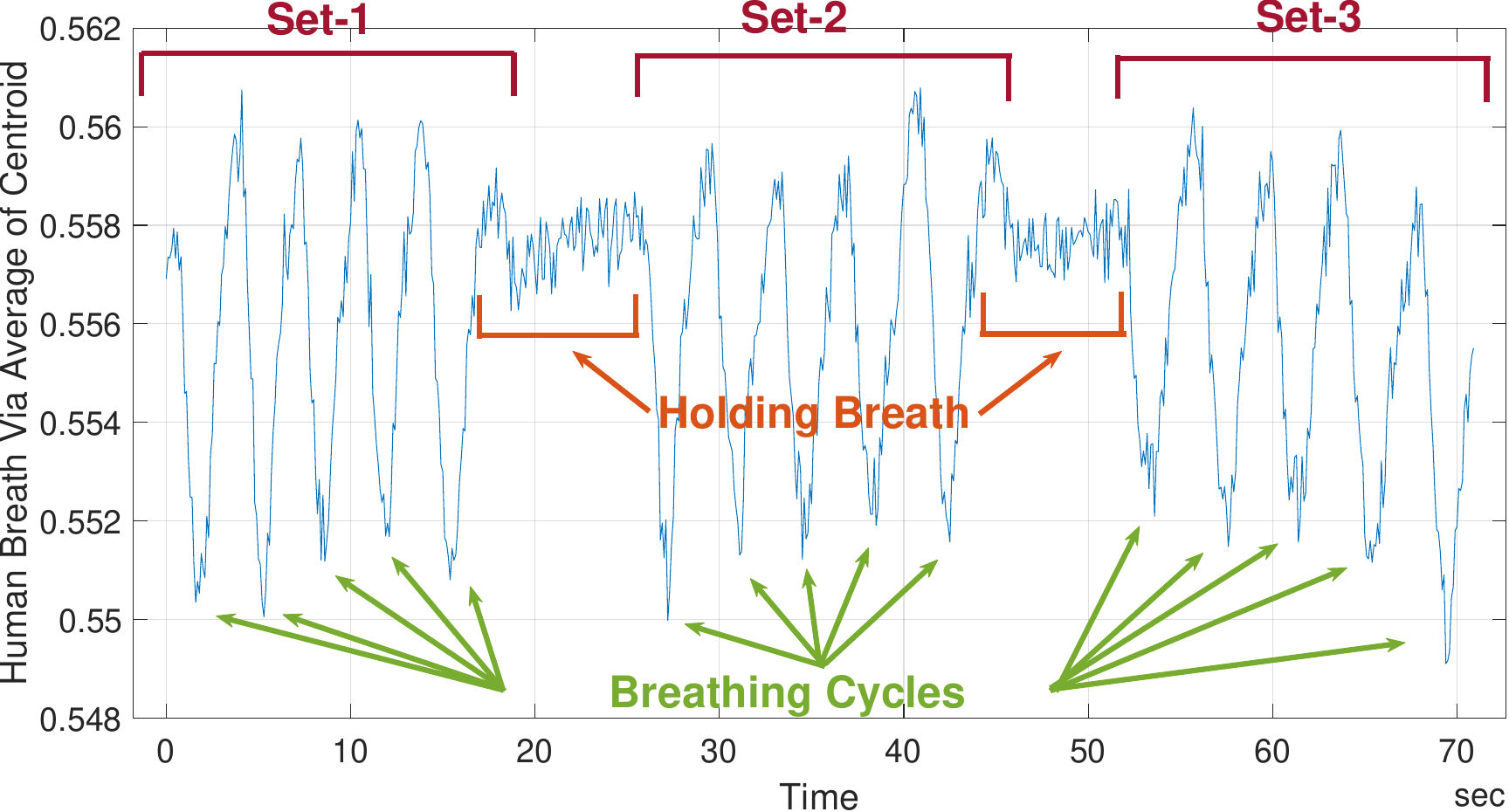}
         \vspace{-3 mm}
         \caption{Illustration of the processed collected data.}
         \label{Fig2}
         \vspace{-7 mm}
\end{figure*}

\vspace{-2 mm}
\section{System Scenario and Experimental Setup} \label{Scenario}
\vspace{-1 mm}
In this study, our objective is to explore the potential of the \ac{LiDAR} sensor in capturing vital breathing moments from various angles. We crafted five distinct experimental scenarios to see how the sensor performs under different conditions, all of which are illustrated in Fig. \ref{Fig1}.

In Scenario-a, the participant was seated upright with the LiDAR sensor directly in front, focusing on frontal torso data. Scenario-b also had the participant seated, but with the LiDAR behind them to assess differences in data from this angle. For Scenario-c and Scenario-d, the LiDAR was placed to the right and left sides, respectively, to examine lateral breathing patterns. Scenario-e involved the participant lying supine with the LiDAR positioned overhead, offering a unique overhead perspective. All seated scenarios used a backless chair to maintain clear line-of-sight between the participant and the LiDAR, and the participant wore a T-shirt for consistency. The LiDAR was kept at a fixed distance of $2$ meters from the participant in every setup. In our experimental procedure, the participant performs three sets of breathing cycles, each set comprising five breaths\footnote{A 'breath' is defined as one inhalation followed by one exhalation, so each set involves five successive inhale-exhale cycles.} After completing each set, the participant pauses and holds breath for a designated period, simulating instances when breathing stops. The duration of each scenario is approximately $1.25$ minutes, and each breath-holding phase spans about $10$ seconds.

Following each experiment, we carefully gather and securely store the data for detailed analysis. Fig. \ref{Fig2} illustrates a sample of the processed data from Scenario-a, displayed over time. A comprehensive explanation of our data processing techniques is discussed in Section \ref{meth}. In Fig. \ref{Fig2}, we identify three distinct sets of breathing patterns, marked in red. Between these sets, periods of breath-holding are indicated in orange. Upon closer inspection, each set contains five individual breathing cycles, each denoted by green arrows.

\vspace{-1 mm}
\section{Methodology} \label{meth}
\vspace{-1 mm}
Utilizing the data gathered from the five scenarios, our aim is to process this data to extract essential breathing parameters. Our objectives are fourfold: $(i)$ continuously tracking inhalation/exhalation patterns over time, $(ii)$ estimating the respiratory rate, $(iii)$ measuring the amplitude or depth of breaths, and $(iv)$ detect any pauses or gaps in breathing.

\vspace{-1 mm}
\subsection{Pre-processing}
\vspace{-1 mm}
The raw \ac{LiDAR} collected data consists of $T$ time frames, such that the frame $n_t\in\mathcal{N}$ where $\mathcal{N}=\{n_1, n_2, \dots, n_t, \dots, n_T\}$ consists of $I$ points where each point $p_i[n_t]$ in the point cloud is defined using its coordinates such that $p_i[n_t]=\left[x_i[n_t], \; y_i[n_t], \;z_i[n_t]\right]^T$. 

We start our analysis by conducting some pre-processing on the collected \ac{LiDAR} raw data to convert it into a format like the one shown in Fig. \ref{Fig2}. In this new format, the data is plotted against time such that each peak or notch in the graph corresponds to a specific breath cycle, which includes both inhalation and exhalation. Once we achieve this format, we will process the newly formatted data so that it is possible to accurately identify each breathing metric.

\begin{figure}[b!]
         \centering
         \includegraphics[width=2.4in]{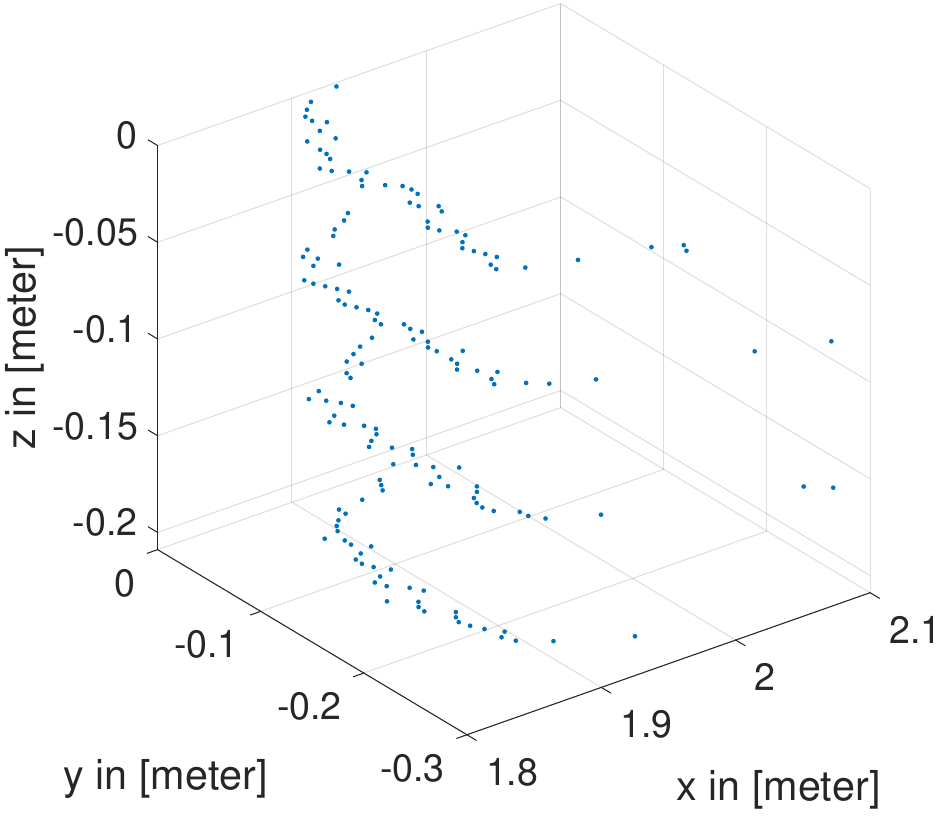}
         \vspace{-2 mm}
         \caption{Human torso point cloud after
         filtering the raw data.}
         \label{Fig3}
\end{figure}

The initial step in our process involves data filtration, where we aim to isolate only the relevant point cloud data associated with the human's torso. Essentially, we want to extract the points related to the torso area and discard all others. An example of this procedure, applied to the data from Scenario-a, can be seen in Fig. \ref{Fig3}. This illustration focuses solely on the participant's chest region. Different techniques can be utilized to to achieve this targeted filtration; for instance, we have demonstrated in \cite{2023_Rinchi_B} a deep-learning approach to segment human body parts utilizing a stand-alone 3D \ac{LiDAR} sensor. Leveraging this technique, we can discern various body sections, thereby segmenting the torso region. The authors in \cite{2022_Bin} address this issue by employing reflective markers worn by humans. Reflective markers can be used to identify torso region. In this work, we utilize a \ac{ROI}-based approach where we define a boundary region such that any point cloud outside this region is filtered. More specifically, let $\mathcal{P}[n_t]$ be the set of all $I$ point cloud data captured by the \ac{LiDAR} at frame $n_t$ (i.e., $\mathcal{P}[n_t]=\{p_{1}[n_t], p_{2}[n_t], \dots, p_{i}[n_t], \dots, p_{I}[n_t]\}, \; \forall n_t \in \mathcal{N}$). Let $\mathcal{T}[n_t]$ represent the defined spatial boundaries of the human torso, such that any point $p_{i}[n_t]\in\mathcal{P}[n_t]$ with coordinates $(x_{i}[n_t],y_{i}[n_t],z_{i}[n_T])$ within $\mathcal{T}[n_t]$ belongs to the chest or abdominal regions. The region $\mathcal{T}[n_t]$ can be defined based on threshold values as
    \vspace{-3 mm}
    \begin{align}
         x_{th1} \le x_{i}[n_t]& \le x_{th2}; \;\;y_{th1} \le y_{i}[n_t] \le y_{th2}; \nonumber \\
        & z_{th1} \le z_{i}[n_t] \le z_{th2},
    \end{align}
{where $x_{th1}, x_{th2}, y_{th1}, y_{th2}, z_{th1}, \text{ and } z_{th2}$ are the thresholds that defines the \ac{ROI}. Then, the filtered point cloud data $\mathcal{F}[n_t]$ in frame $n_t$ are given by}
\vspace{-2.5 mm}
\begin{equation}
   \mathcal{F}[n_t]=\mathcal{P}[n_t] \cap \mathcal{T}[n_t],
   \vspace{-2 mm}
\end{equation}
where the new filtered data set $\mathcal{F}[n_t]$ consists of $E$ points, i.e., $\mathcal{F}[n_t]=\{p^F_1[n_t], p^F_2[n_t], \dots, p^F_e[n_t], \dots, p^F_E[n_t]\}$ such that $p^F_e[n_t]=[x_e^F[n_t], \; y_e^F[n_t], \; z_e^F[n_t]]^T.$
The boundary set $\mathcal{T}[n_t]$ can be configured in one of two ways: it can remain stationary, reflecting empirical domain knowledge (e.g., when a human lies on a bed with the LiDAR positioned above the torso), or it can be dynamically adjusted using a torso tracking mechanism \cite{2023_Cherif}. In this work, we follow the stationary approach.

\begin{figure}[t!]
         \centering
         \includegraphics[width=2.2in]{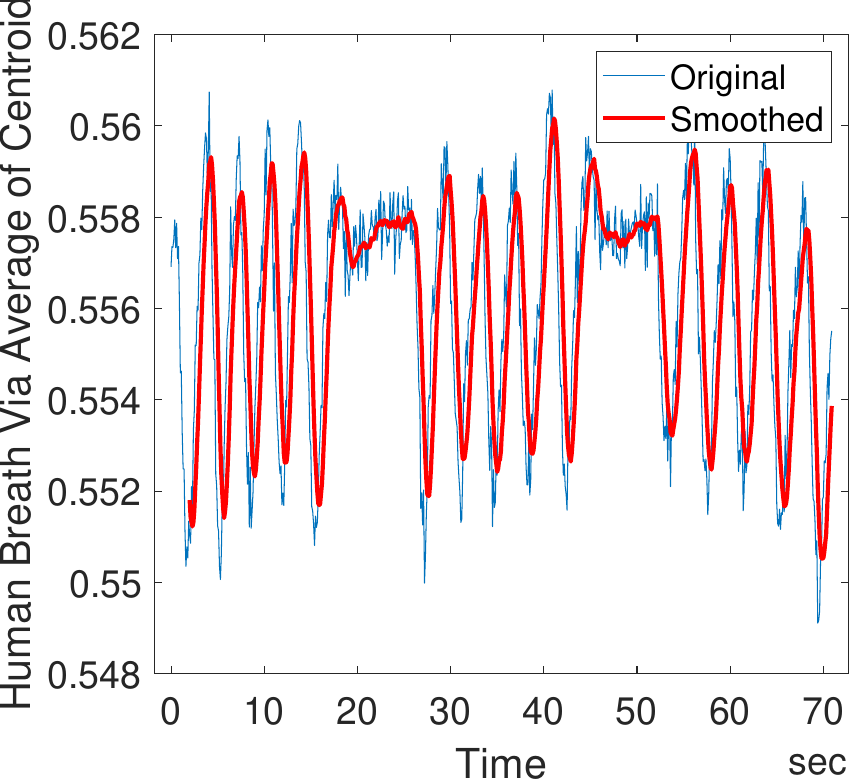}
         \vspace{-3 mm}
         \caption{Illustration of smoothing the noisy data.}
         \label{Fig4}
         \vspace{-7 mm}
\end{figure}

The subsequent step involves converting the data to a format that better captures breathing patterns over time. This involves leveraging the geometric shifts that arise during the breathing process and charting these changes over time. Our approach begins with selecting a representative point, $P^\star[n_t]$, from the filtered human torso dataset $\mathcal{F}[n_t]$. While there are various methods to pinpoint this representative point, we opt for the centroid of the filtered human torso dataset. Our preference for the centroid is grounded in its robustness to outliers, ensuring a more accurate representation even if the data filtering isn't flawless. Accordingly, we define the representative point, $P^\star[n_t]$ as
\vspace{-3 mm}
\begin{align}
    P^\star[n_t]=& \{C^{x}[n_t],C^{y}[n_t],C^{z}[n_t]\}, \;\;\;\;\text{where} \nonumber \\
    C^{x}[n_t] &\!=\! E^{-1} \sum_{e=1}^{E}x_{e}^F[n_t]; \; C^y[n_t] = E^{-1} \sum_{e=1}^{E}y_{e}^F[n_t]; \nonumber \\
    \; C^z[n_t] &= E^{-1} \sum_{e=1}^{E}z_{e}^F[n_t].
\end{align}

Now, it is possible to capture human breaths by inspecting the changes in the values of $C^{x}[n_t],C^{y}[n_t],C^{z}[n_t]$. To visualize these variations, one approach is to plot specific coordinate components of $P^\star[n_t]$ against time — for instance, plotting $C^x[n_t]$ versus the corresponding frames $n_t\in\mathcal{N}$. Additionally, an alternative method involves computing the mean value of the three centroid coordinates and charting this average against time, represented as $\frac{C^x[n_t]+C^y[n_t]+C^z[n_t]}{3}$ versus the corresponding frames $n_t\in\mathcal{N}$. The choice of the utilized approach depends on the orientation of the human with respect to the \ac{LiDAR} sensor. Formatting the data against specific coordinate components requires the knowledge of the orientation of the human with respect to the \ac{LiDAR} but can yield higher sensitivity to capture the inhalation/exhalation cycle, while formatting the data against the average of the centroid can ensure a generic solution that is independent of the human orientation but will yield lower sensitivity to capture the breath compared to the former approach.

\vspace{-1 mm}
\subsection{Breath Monitoring}
\vspace{-2 mm}
Based on the established methodology, we can now generate a breath-versus-time plot akin to Fig. \ref{Fig2}. The subsequent task involves identifying human breaths from the presented data. Notably, the peaks in the diagram align with human chest displacements. Yet, as illustrated in Fig. \ref{Fig2}, the inherent noise in the data—attributable to LiDAR's acute sensitivity to chest movements—complicates peak detection. To address this challenge, we introduce a data smoothing technique using a moving average filter. This transforms the data into a more streamlined representation, comparable to the red curve in Fig.~\ref{Fig4}. The mathematical representation for the moving average filter is as follows
\vspace{-3 mm}
\begin{equation} \label{MA}
    y_{\text{smoothed}}[n_t]=1/M\sum_{m=0}^{M-1}y[n_t-m], \; \forall n_t\in \mathcal{N},
    \vspace{-2 mm}
\end{equation}
where $M$ is the length of the moving average window (number of frames to average over), $m$ is the summation index, representing the samples within the window, $y[n_t-m]$ is un-smoothed data at frame $n_t$, and $y_{\text{smoothed}}[n_t]$ is the output smoothed data. The ability to accurately track inhalation and exhalation patterns over time is essential for understanding specific events of the breath cycle. Towards this end, the first step is to detect the peaks within the smoothed breath data, as they correspond to these breath cycle events. For this task, we employed a straightforward yet effective local minimums detection technique. This method is mainly rooted in pinpointing data points that exhibit values higher than their immediate neighbors.

However, to further refine our results and improve accuracy, we introduced an additional filtering step. Instead of relying exclusively on the initial set of detected peaks, we filter out local minimums (given that in our data, breath events manifest as negative peaks) that fall above the average value of the smoothed data. The underlying rationale is that notable breath events are typically characterized by pronounced troughs, which are deeper than the average of the entire data set. Considering that $\mathcal{B}=\{{b}[{\tilde{n}_1}], {b}[{\tilde{n}_2}], \dots, {b}[{\tilde{n}_k}], \dots {b}[{\tilde{n}_K}]\}$ is a set comprising the amplitudes of the detected peaks from the smoothed data and $\tilde{\mathcal{N}}=\{\tilde{n}_1, \tilde{n}_2, \dots, \tilde{n}_k, \dots \tilde{n}_K\}$ is a set containing their corresponding frame numbers, the filtered set $\bar{\mathcal{B}}=\{\bar{b}[\bar{n}_1], \bar{b}[\bar{n}_2], \dots, \bar{b}[\bar{n}_s], \dots \bar{b}[\bar{n}_S]\}$ and its respective frame set $\bar{\mathcal{N}}=\{\bar{n}_1, \bar{n}_2, \dots, \bar{n}_s, \dots \bar{n}_S\}$ can be defined as
\vspace{-3 mm}
\begin{align}
  \bar{\mathcal{B}}\!&=\!\!\{{b}[\tilde{n}_t] \;|\; {b}[\tilde{n}_t] \!\le \! \frac{1}{T}\sum_{n_t=1}^T y_{\text{smoothed}}[n_t], \forall \tilde{n}_t \in \tilde{\mathcal{N}}, n_t \in \mathcal{N}\}, \nonumber \\
  \bar{\mathcal{N}}\!&=\!\!\{\tilde{n}_t \;|\; {b}[\tilde{n}_t] \!\le \!\frac{1}{T}\sum_{n_t=1}^T y_{\text{smoothed}}[n_t], \forall \tilde{n}_t \in \tilde{\mathcal{N}}, n_t \in \mathcal{N}\}.
\end{align}
Accordingly, the breathing patterns can be tracked using $\bar{\mathcal{N}}$ while we account for $\bar{\mathcal{B}}$ to monitor the breathing depth.

Following the identification of breath cycle events, the next imperative step is to quantify the respiratory rate, a measure which plays a pivotal role in many clinical and physiological assessments. The respiratory rate, denoted as $R$, represents the number of complete inhalation-exhalation cycles an individual completes in a minute. Given the detected breathing events denoted by the set $\bar{\mathcal{B}}$, the respiratory rate can thus be estimated as
\vspace{-2 mm}
\begin{equation} \label{RR}
R=\frac{\text{card}\left(\bar{\mathcal{B}}\right)}{\tau}\times 60,
\vspace{-2 mm}
\end{equation}
where $\text{card}\left(\bar{.}\right)$ refers to the cardinality (or count) of a set (i.e., the number of elements in a set), and $\tau$ is measurement time in seconds. The last breathing metric that is to be identified is the episodes of breathlessness. We define this metric as the set that includes the frames in which a human was not breathing, i.e., $\widehat{\mathcal{N}}=\{\widehat{n}_1, \widehat{n}_2, \dots, \widehat{n}_g, \widehat{n}_G\}$  where $\widehat{\mathcal{N}}\in {\mathcal{N}}$ and $\bar{\mathcal{N}} \cap \widehat{\mathcal{N}}=\varnothing$. For this purpose, we estimate $\widehat{\mathcal{N}}$ using a moving variance, more specifically, we start by computing the moving variance for every point as
\vspace{-2 mm}
\begin{equation} \label{MV}
    v[n_t]=\frac{1}{W-1}\sum_{u=n_t-\lfloor{\frac{W-1}{2}}\rfloor}^{{n_t+\lfloor{\frac{W}{2}}}\rfloor}\left( y_{\text{smoothed}}[u]-\mu[n_t]\right)^2, \; \forall n_t\in\mathcal{N},
    \vspace{-1 mm}
\end{equation}
where $W$ is moving variance window size, $\lfloor.\rfloor$ is the floor function which round down to the nearest integer, and $\mu[n_t]$ is the mean of the data points in the window for point $n_t$ which can be defined as
\vspace{-2 mm}
\begin{equation}
    \mu[n_t]=\frac{1}{W}\sum_{u=n_t-\lfloor{\frac{W-1}{2}}\rfloor}^{{n_t+\lfloor{\frac{W}{2}}}\rfloor} y_{\text{smoothed}}[u], \; \forall n_t \in \mathcal{N}.
    \vspace{-3 mm}
\end{equation}
Now we use the calculated moving average to estimate $\widehat{\mathcal{N}}$ using a threshold value $\gamma$ as 
\vspace{-2 mm}
\begin{equation} \label{NB}
  \widehat{\mathcal{N}}=\{n_t \;| \; {v}[n_t] \le \gamma, \forall n_t \in \mathcal{N}\}.
  \vspace{-3 mm}
\end{equation}

\begin{figure}[t!]
        \vspace{2 mm}
         \centering
         \includegraphics[width=2.5in]{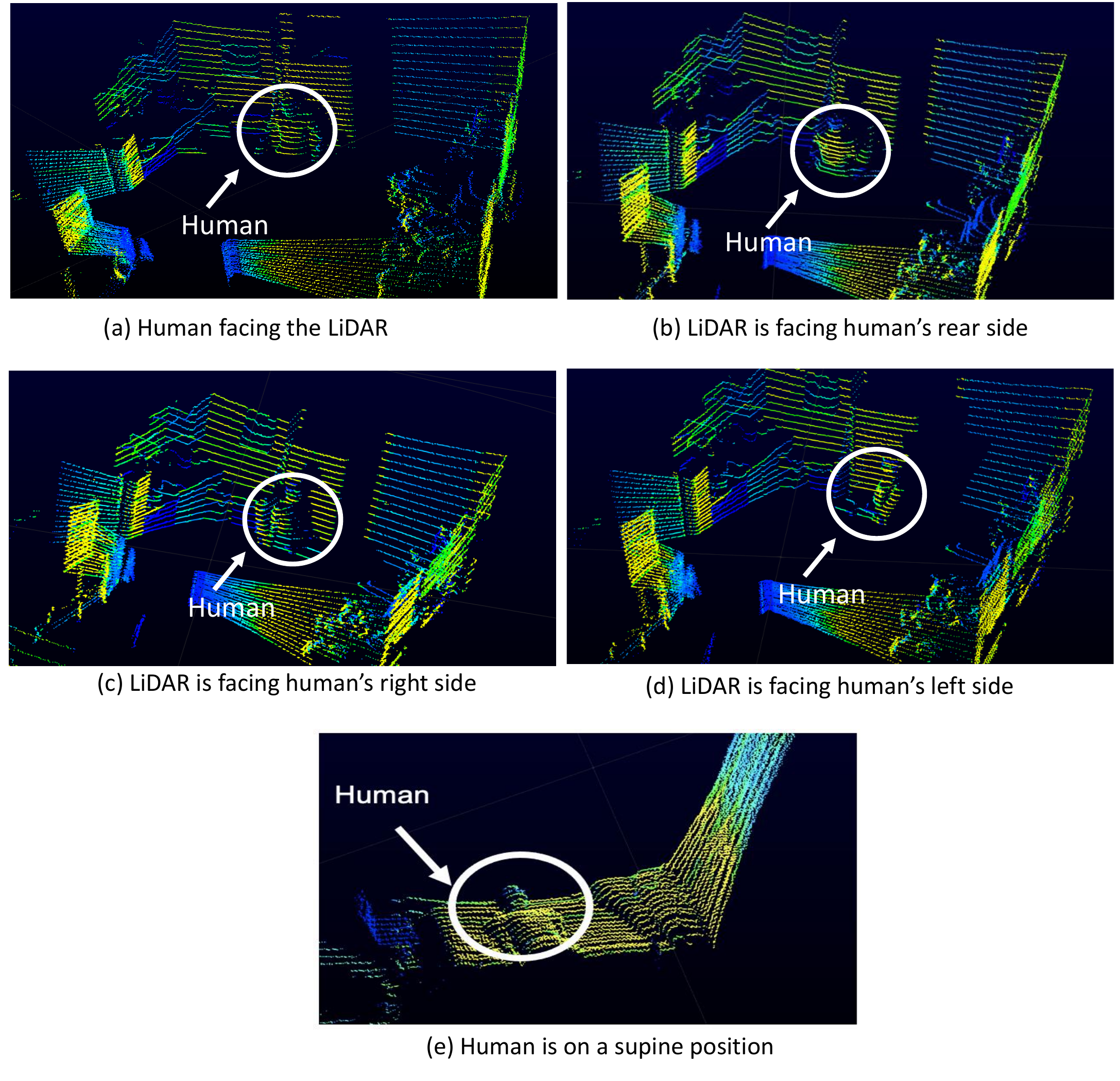}
         \vspace{-2 mm}
         \caption{Illustration of the conducted experimentation.}
         \label{Fig5}
         \vspace{-7 mm}
\end{figure}
\vspace{-2 mm}
\section{Numerical Results}
\vspace{-2 mm}
In this section, we present selected numerical results. As stated in Section. \ref{Scenario}, we evaluate the performance of the methodology conducted for five distinct scenarios, a snapshot of the raw collected data of these five scenarios is shown in Fig. \ref{Fig5}, while we utilized the Velodyne PUCK \ac{LiDAR} sensor to conduct all these measurements. We experimentally specified stationary ROI thresholds $x_{th1}, \; x_{th2}, \; y_{th1}, \; y_{th2}, \; z_{th1}, \; \text{and} \; z_{th2}$ in a scenario-specific manner. The number of points per raw point cloud collected $I$, the number of the filtered point cloud $E$, and the number of time frames $T$ are all dependent values based on the \ac{LiDAR} measurement in every distinct scenario. The window size of the moving average window $M$ in \eqref{MA} is set to $10$, while the window size of the moving variance window $W$ in \eqref{MV} is set to be $25$. The threshold value for identifying the holding breath frames $\gamma$ in \eqref{NB} is set to $1\times 10^{-6}$. Finally, this study was institutional review board (IRB) approved via University of Missouri human subjects research office. 

As we collect four metrics in our methodology and that are $(i)$ the frames of breathing set denoted by the set $\bar{\mathcal{N}}$, $(ii)$ the frames of holding breath set denoted by the set  $\widehat{\mathcal{N}}$, $(iii)$ the breathing depth denoted by the set  $\bar{\mathcal{B}}$, and $(iv)$ the respiratory rate denoted by $R$; we use \textit{Accuracy} to evaluate the first two metrics, while we use the \textit{Root Mean Square Error (RMSE)} to evaluate the latter two. We define \textit{Accuracy} as
\vspace{-2 mm}
\begin{equation}
    Accuracy=\frac{TP+TN}{TP+TN+FP+FN}\, ,
    \vspace{-2 mm}
\end{equation}
where $TP$ refers to the true positive, $TN$ refers to the true negative, $FP$ refers to the false positive, and $FN$ refers to the false negative. The \textit{RMSE} is defined as
\vspace{-2 mm}
\begin{equation}
\textit{RMSE}=\sqrt{\frac{1}{K}\sum_{k=1}^K\left(y_k-\widehat{y}_k\right)^2},
\vspace{-2 mm}
\end{equation}
where $y_k$ is the estimated value, and $\widehat{y}_k$ is the ground truth. For both metrics, we rely on manual ground truth constitution in which we manually specify the frames of breath, the frames of holding breath, the breath depth, and the respiratory rate based on our inspection of the data.

\begin{table}[t]
\vspace{3 mm}
\caption{{Breathing monitoring numerical results}}
\vspace{-3 mm}
\centering
\label{results}
\begin{tabular}{lllll}
\hline
Scenario   & \begin{tabular}[c]{@{}l@{}}Breathing \\ \textit{Accuracy}\end{tabular} & \begin{tabular}[c]{@{}l@{}}Holding Breath \\ \textit{Accuracy}\end{tabular} & \begin{tabular}[c]{@{}l@{}}Breath \\ Depth \textit{RMSE}\end{tabular} & \begin{tabular}[c]{@{}l@{}}Respiratory \\ rate \textit{RMSE}\end{tabular} \\ \hline
Scenario-a & 1.00                                                          & 0.93                                                              & 0.0019                                                      & 0.00                                                            \\
Scenario-b & 0.73                                                          & 0.85                                                              & 0.0020                                                      & 3.21                                                            \\
Scenario-c & 0.87                                                          & 0.89                                                              & 0.0014                                                      & 1.63                                                            \\
Scenario-d & 0.93                                                          & 0.92                                                              & 0.0015                                                      & 0.78                                                            \\
Scenario-e & 1.00                                                          & 0.94                                                              & 0.0014                                                      & 0.00                                                            \\ \hline
\end{tabular}
\vspace{-8 mm}
\end{table}

The experimental results are delineated in Table. \ref{results}, which highlights the dependence of breathing performance metrics on the spatial alignment of the human subject with respect to the \ac{LiDAR} sensor. Optimal breathing detection accuracy is reported for scenarios where the LiDAR is aligned with the front of the human subject (Scenario-a and Scenario-e). The rear-facing orientation (Scenario-b) resulted in the least accurate breathing performance, whereas side-facing orientations (Scenario-c and Scenario-d) delivered intermediate accuracy. This variation in performance can be attributed to the underlying detection mechanism, which is predicated on capturing the thoracic movements induced by respiration, movements that are more prominently manifested and hence more readily detected at the front of the torso than at the rear.

The RMSE associated with respiratory rate reflects the critical dependence on the detection precision of individual breath cycles. Misidentification or omission of a cycle can significantly skew the respiratory rate estimation, as articulated in \eqref{RR}. For the depth of breathing, the RMSE was calculated exclusively for correctly identified breathing cycles. This selective evaluation has yielded a lower RMSE for breathing depth, indicating a high fidelity in quantifying the amplitude of thoracic movements during respiration for each correctly detected cycle. It is important to note that while depth estimation is robust in the various scenarios, the overall reliability of respiratory monitoring is dependent on the accuracy of breath cycle detection.

\begin{figure}[t!]
         \centering
         \includegraphics[width=2.7in]{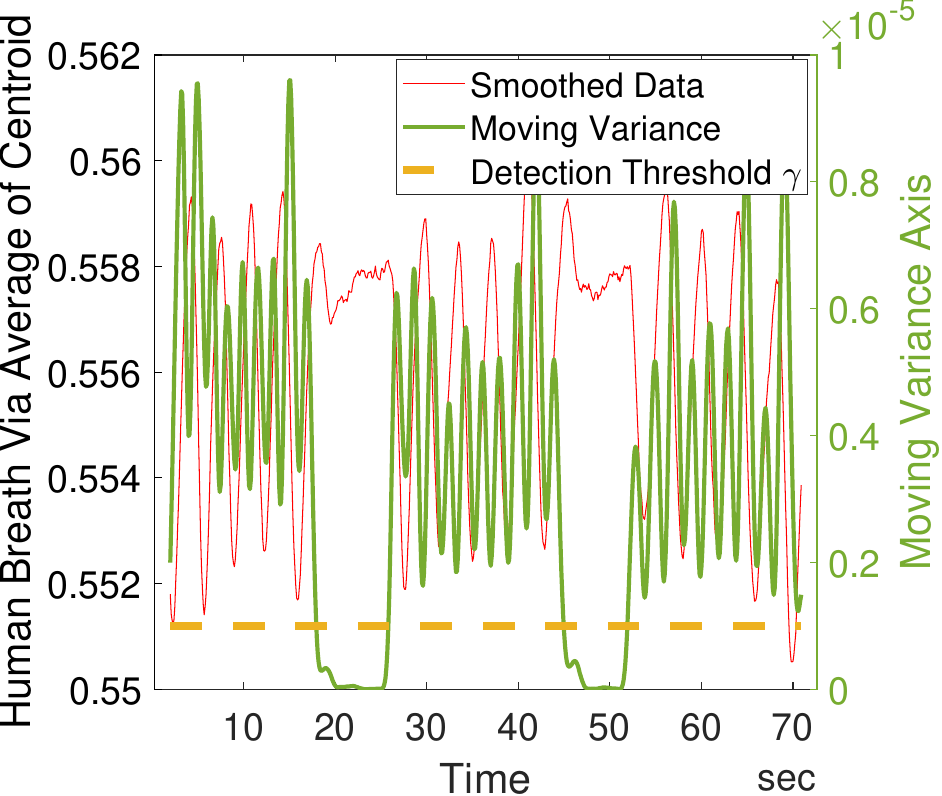}
         \vspace{-2 mm}
         \caption{Illustration of holding breath detection mechanism.}
         \label{Fig6}
         \vspace{-5 mm}
\end{figure}

Fig. \ref{Fig6} showcases the process of identifying breath-holding events in Scenario-a, following the methods outlined in \eqref{MV} and \eqref{NB}. The red curve illustrates the smoothed data, and the green curve depicts its moving variance, indicating chest movement variability over time. A drop in moving variance below the threshold $\gamma$ (the yellow dashed line) signals minimal chest movement, pointing to breath-holding. Despite the threshold $\gamma$'s sensitivity, as detailed in Table. \ref{results}, this method effectively detects respiratory events, enabling real-time monitoring and swift responses in urgent care. This precision highlights our system's reliability and its value in clinical environments where accurate respiratory tracking is crucial.

\vspace{-2 mm}
\section{Conclusion}
\vspace{-2 mm}
This study explores \ac{LiDAR} technology's effectiveness in remote breathing monitoring by converting torso movements into a signal that highlights various breathing metrics. Through experiments in five unique scenarios involving different subject orientations to the \ac{LiDAR} sensor, we found that while subject orientation affects measurement accuracy, \ac{LiDAR} reliably monitors respiration across all situations. This demonstrates \ac{LiDAR}'s robustness and adaptability for real-world use, where controlling subject-sensor positioning is challenging, confirming its potential for versatile respiratory monitoring applications.

\bibliographystyle{IEEEtran}
\bibliography{referances.bib}

\end{document}